\documentclass[useAMS,usenatbib]{mn2e}
 
\usepackage{graphicx}
\usepackage{natbib}
\usepackage{txfonts}
\usepackage{color}
\usepackage{graphicx}
\usepackage{multirow}

\def\lta{\lower2pt\hbox{$\buildrel {\scriptstyle <} 
   \over {\scriptstyle\sim}$}}
\def\gta{\lower2pt\hbox{$\buildrel {\scriptstyle >} 
   \over {\scriptstyle\sim}$}}
\newcommand{\fracb}[2]{\left(\frac{#1}{#2}\right)}

\title[Scaling relations in numerical simulations]
{Scaling relations between numerical simulations and physical systems they represent}

\author[J. Granot]{Jonathan Granot$^{1,2,3}$\\
$^{1}$Racah Institute of Physics, The Hebrew University, Jerusalem 91904, Israel\\
$^{2}$Raymond and Beverly Sackler School of Physics \& Astronomy, Tel Aviv University, Tel Aviv 69978, Israel\\
$^{3}$Centre for Astrophysics Research, University of Hertfordshire, College Lane, Hatfield, AL10 9AB, UK}

\begin{document}
 

\pagerange{\pageref{firstpage}--\pageref{lastpage}} \pubyear{2011}

\label{firstpage}

\maketitle

\begin{abstract}
The dynamical equations describing the evolution of a physical system
generally have a freedom in the choice of units, where different
choices correspond to different physical systems that are described by
the same equations. Since there are three basic physical units, of
mass, length and time, there are up to three free parameters in such a
rescaling of the units, $N_f\leq{}3$. In Newtonian hydrodynamics,
e.g., there are indeed usually three free parameters, $N_f=3$. If,
however, the dynamical equations contain a universal dimensional
constant, such as the speed of light in vacuum $c$ or the
gravitational constant $G$, then the requirement that its value
remains the same imposes a constraint on the rescaling, which reduces
its number of free parameters by one, to $N_f = 2$. This is the case,
for example, in magneto-hydrodynamics or special-relativistic
hydrodynamics, where $c$ appears in the dynamical equations and forces
the length and time units to scale by the same factor, or in Newtonian
gravity where the gravitational constant $G$ appears in the
equations. More generally, when there are $N_{\rm udc}$ independent
(in terms of their units) universal dimensional constants, then the
number of free parameters is $N_f=\max(0,3-N_{\rm udc})$. When both
gravity and relativity are included, there is only one free parameter
($N_f = 1$, as both $G$ and $c$ appear in the equations so that
$N_{\rm udc}=2$), and the units of mass, length and time must all
scale by the same factor. The explicit rescalings for different types
of systems are discussed and summarized here. Such rescalings of the
units also hold for discrete particles, e.g. in $N$-body or particle
in cell simulations. They are very useful when numerically
investigating a large parameter space or when attempting to fit
particular experimental results, by significantly reducing the
required number of simulations.
\end{abstract}


\section{\bf Introduction}

Numerical simulations are gradually but steadily playing an
increasingly more important role in the study of many physical
systems.  In particular, they have a growing impact on many different
fields, such as fluid dynamics, plasma physics, astrophysics, particle
physics, Earth and planetary sciences, and meteorology. In particular,
numerical computation is a vital tool for studying complex problems
that are hard to solve analytically, such as nonlinear, many-body or
multiple scale processes. 
 
Here I outline how for many types of numerical simulations, the
results of a single simulation correspond to a whole family of
physical systems. This arises from the more general freedom in the
choice of units in the dynamical equations that describe a particular
type of physical system. This freedom (or lack thereof) also holds for
solutions of these equations, as long as the initial or boundary
conditions do not impose additional universal dimensional constants
(UDCs), thus increasing $N_{\rm udc}$. It is equally valid for
analytic or numerical solutions of these equations. Here the focus is
on numerical solutions, and in particular on numerical simulations.

In this work it is assumed that the number of basic physical units is
$N_{\rm bpu} = 3$, corresponding to units of mass ($m$), length ($l$)
and time ($t$). This means that electric charge is expressed in terms
of these units ($[q] = m^{1/2}l^{3/2}t^{-1}$), which corresponds to
c.g.s Gaussian units in which Maxwell's equations contain one UDC --
the speed of light in vacuum, $c$. The number of free parameters $N_f$
that span the family of physical systems that correspond to a single
simulation is generally given by
\begin{equation}\label{eq:N_f_gen}
N_{\rm f} = \max\left(0,\,N_{\rm bpu} - N_{\rm udc}\right)\ .
\end{equation}
For magneto-hydrodynamics (MHD), either Newtonian or special
relativistic, where the only UDC is $c$ (i.e. $N_{\rm udc} = 1$), this
implies that $N_f = 3-1 = 2$. 

An alternative choice of basic physical units is SI units, in which
electric charge is explicitly treated as a fourth basic physical unit
(measured in Coulombs), so that $N_{\rm bpu} = 4$. In these units,
however, Maxwell's equations contain two UDCs, $\epsilon_0$ (the
permittivity of free space or the electric constant) and $\mu_0$ (the
permeability of free space or the magnetic constant), where
$(\epsilon_0\mu_0)^{-1/2} = c$. Therefore, with this choice of units
$N_{\rm udc} = 2$ for MHD, leading to $N_f = 4-2 = 2$, i.e. the same
number of free parameters as before.  This nicely demonstrates that
the freedom in rescaling the basic physical units for a given physical
system is independent of the specific choice of basic
units. Therefore, the particular choice that we use in this work
should not affect $N_f$ and the implied relations between the allowed
scalings of the units of mass, length and time.

The concept of UDCs is not new. For example, \citet{Ellis68} has
referred to UDCs as universal scale-dependent constants, and
demonstrated that if units of mass and force are expressed in terms of
length and time ($[m]\to l^3t^{-2}$, $[F]\to l^4t^{-4}$) then this
renders the gravitational constant $G$ dimensionless in such units, so
that an appropriate choice of their magnitude can make it equal to
unity and thus disappear from Newton's law of gravity. In such units,
$N_{\rm udc} = 2$ (as the basic physical units are only of length and
time) while $N_{\rm udc} = 0$ for Newtonian gravity, leading to $N_f =
2-0=2$. For the regular units $N_{\rm bpu} = 3$ and $N_{\rm udc} = 1$
for Newtonian gravity, again leading to $N_f = 3-1=2$. Unlike the
c.g.s Gaussian units that were constructed in order to eliminate the
constant [or SI UDC $1/(4\pi\epsilon_0)$] in Coulomb's law, the above
choice of units that eliminate $G$ from Newton's law of gravity do not
work very well in general. The reason for this is the equivalence
(which does not have an electromagnetic analog) between gravitational
mass, for which these units were constructed, and inertial mass, which
appears also in systems in which gravity is unimportant and can be
neglected. In such systems that choice of units will artificially
eliminate a degree of freedom in the rescaling of the units for no
good reason.

In \S~\ref{sec:sim} the scalings are explicitly derived for different
types of simulations, and summarized in Table~\ref{tab:sum}. Some
caveats, namely the possible increase in $N_{\rm udc}$ for certain
equations of state or radiative processes, are discussed in
\S~\ref{sec:caveats}. In \S~\ref{sec:afterlow} the application of
special relativistic hydrodynamic simulations to model gamma-ray burst
(GRB) afterglow is discussed in some more detail, as a useful case
study. The conclusions are discussed in \S~\ref{sec:dis}.

\section{\bf Scaling relations for different types of numerical simulations}
\label{sec:sim}

Physical systems are usually described either in the continuum limit,
such as in hydrodynamics or magneto-hydrodynamics (MHD), or by
following the motions of discrete particles. Numerically, the former
description corresponds to hydrodynamic
\citep[e.g.,][]{Li10,LB11,Wallace11} or MHD
\citep[e.g.,][]{FM08,Ishihara09,Amit10} simulations, while the latter
includes examples such as cosmological $N$-body simulations
\citep[with a large number $N\gg 1$ point particles that interact only
through their mutual gravitational attraction;
e.g.][]{KKK97,NFW97,Springel05} or particle in cell (PIC) plasma
simulations \citep[where a large number of point particles with both
positive and negative electric charges interact electromagnetically;
e.g.][]{Birn01,PM02,Spit08}. Table~\ref{tab:sum} summarizes the
allowed scalings for different types of simulations. Hybrid
simulations that include both a continuous medium and discrete
particles are also possible
\citep[e.g.,][]{KWH96,Gnedin04,Spr05}, and the restrictions on them
can be derived in a straightforward manner by combining the
restrictions on their constituents. This is manifested in the total
number of independent (in terms of their units) UDCs, $N_{\rm udc}$,
which determines the number of free parameters $N_f$ that span the
family of physical systems that correspond to a single simulation
through
\begin{equation}\label{eq:N_f}
N_{\rm f} = \max\left(0,\,3 - N_{\rm udc}\right)\ ,
\end{equation}
where we assume for the rest of this work that there are three basic
physical units ($N_{\rm bpu} = 3$) of mass, length and time.

\vspace{0.2cm}
\noindent{\bf Hydrodynamic Simulations:}
here we outline how the results of different types of hydrodynamic
simulations correspond to a family of physical systems. The numerical
code solves, e.g., for the proper rest-mass density $\rho$, pressure
$p$, and velocity $\vec{v} = \vec{\beta} c$, as a function of time and
space, $(t,\vec{r})$, and assumes some equation of state that relates
the specific enthalpy to the pressure and density. The evolution of
these quantities is usually solved over a finite volume $V$ and time
range $t_i \leq t \leq t_f$. The initial conditions at $t_i$ must be
specified over the volume $V$, as well as boundary conditions at the
edges of this volume at $t_i \leq t \leq t_f$.  In order to solve the
hydrodynamic equations numerically, they are first made dimensionless
by moving to code units that are determined by choosing some specific
scales (which we shall denote by a subscript ``0'') for the three
basic physical units of mass ($m_0$), length ($l_0$), and time
($t_0$). The corresponding dimensionless variables in code units are
denoted by a twiddle, where a general quantity $Q$ with units of
$m^{A}l^{B}t^{C}$ corresponds to $\tilde{Q} = Q/(m_0^A l_0^B t_0^C)$.

Once a particular initial physical configuration is mapped onto the
dimensionless code variables, the hydrodynamic equations are solved
for these variables. Then, the numerical solution is usually
translated back to the original physical units, $Q = \tilde{Q}\,m_0^A
l_0^B t_0^C$. This is not a unique procedure, however, because of the
freedom in the choice of units that was described above. Therefore,
the same numerical solution also holds equally well for a whole set or
family of different physical systems, which correspond to different
choices for the basic physical units, $(m'_0,\,l'_0,\,t'_0) = (\zeta
m_0,\, \alpha l_0,\, \eta t_0)$.  Such different choices of units can
conveniently be implemented when switching back from the dimensionless
code variables to the corresponding variables with physical units.

For a purely Newtonian simulation without gravity, there are
indeed three free parameters ($\zeta$, $\alpha$ and $\eta$),
i.e. $N_f = 3$ since there are no relevant UDCs ($N_{\rm udc} = 0$),
and the family of physical systems represented by the simulation is
given by
\begin{eqnarray}\nonumber
Q' = \tilde{Q}\,(m'_0)^A (l'_0)^B (t'_0)^C = 
\zeta^A\alpha^B\eta^C\tilde{Q}\,m_0^A l_0^B t_0^C = \zeta^A\alpha^B\eta^C Q
\\ \label{eq:scaling}
\Longleftrightarrow\quad 
Q'\left(t' = \eta t,\,\vec{r'} = \alpha\vec{r}\right) =
\zeta^A\alpha^B\eta^C Q(t,\vec{r})\ .\quad\quad
\end{eqnarray}
The scaling $\vec{r'} = \alpha\vec{r}$ reads for
different coordinate systems,
\begin{eqnarray}\nonumber
(x',\,y',\,z') &=& (\alpha x,\, \alpha y,\, \alpha z)\quad{\rm cartezian}\ ,
\\ \nonumber
(z',\,r'_{\rm cyl},\,\theta') &=& (\alpha z,\, \alpha r_{\rm cyl} ,\, \theta)\quad{\rm cylindrical}\ ,
\\
(r',\,\theta',\,\phi') &=& (\alpha r,\,\theta,\,\phi)\quad{\rm spherical}\ .
\end{eqnarray}
Similarly, the initial conditions would be at $(t',\vec{r'}) = (\eta
t_i,\alpha\vec{r})$ and the boundary conditions would be at the edge
of $V' = \alpha^3 V$ or $(t'_i\leq t'\leq t'_f,\vec{r'}_{\rm edge})$
where $\vec{r'}_{\rm edge} = \alpha\vec{r}_{\rm edge}$, $t'_i = \eta
t_i$ and $t'_f = \eta t_f$.

When there are relativistic velocities, either of bulk motions or
random motions of the particles (i.e. relativistic temperatures), then
we no longer have $\beta \ll 1$ and $p/\rho \ll c^2$, so that
relativistic effects in the dynamical equations (which depend on the
bulk Lorentz factor $\Gamma$) or the equation of state (which depends
on $p/\rho$) can no longer be neglected. This requires
that\footnote{For $\Gamma\gg 1$ there exists a different type of
rescaling that allows $\Gamma'$ to vary relative to $\Gamma$
\citep{MGA09}, but this is not a rescaling of the basic physical
units, and it is valid only for a forward-reverse shock system in
which the forward shock is ultra-relativistic.}  $\Gamma'=\Gamma$
(i.e. $v'=v$) and $p'/\rho' = p/\rho$, where unprimed quantities are
the usual ones for $\zeta = \alpha = \eta = 1$, which implies that
$\eta = \alpha$, i.e. that the length and time units scale by the same
factor. A more elegant way of deriving this is that special relativity
introduces $c$ (the speed of light in vacuum), a UDC with units of
$l/t$, and thus requires $\alpha/\eta = 1$ in order to keep its value
the same in all our family of physical systems. This reduces the
number of free scaling parameters to two ($N_f = 2$ since $N_{\rm udc}
= 1$). In particular, $x'_\mu = (t',\vec{r'}) = \alpha(t,\vec{r}) =
\alpha x_\mu$, $\rho'/\rho = p'/p =\zeta\alpha^{-3}$, and the family
of physical systems corresponding to a particular simulation is given
by
\begin{equation}\label{eq:c}
(m'_0,\,l'_0,\,t'_0) = (\zeta m_0,\, \alpha l_0,\, \alpha t_0)\ ,\quad
Q'(x'_\mu = \alpha x_\mu) = \zeta^A\alpha^{B+C} Q(x_\mu)\ .
\end{equation}
In particular, the total energy $E$, and mass $M$ either in a
particular computational cell or in the whole computational box (or
volume $V$) scale as $E'/E = M'/M = \zeta$.

When gravity is included, this introduces the gravitational
constant $G$ -- a UDC with units of $m^{-1}l^3t^{-2}$. Thus, in order
for it to keep the same value in all our family of systems requires
that
\begin{equation}\label{eq:gravity}
(l')^3(m')^{-1}(t')^{-2} = l^3m^{-1}t^{-2}
\quad\Longleftrightarrow\quad \zeta = \alpha^3\eta^{-2}\ .
\end{equation}

When there is only weak or Newtonian gravity (and general
relativistic effects can be neglected), and Newtonian (bulk or
thermal) motions, then $G$ is the only UDC ($N_{\rm udc} = 1$, $N_f =
2$) and
\begin{eqnarray}\nonumber
(m'_0,\,l'_0,\,t'_0) = (\alpha^3\eta^{-2} m_0,\, \alpha l_0,\, \eta t_0)\ ,\quad
\\ \label{eq:G}
Q'\left(t' = \eta t,\,\vec{r'} = \alpha\vec{r}\right) =
\alpha^{B+3A}\eta^{C-2A} Q(t,\vec{r})\ .
\end{eqnarray}

When there effects of general relativity cannot be ignored, or
for Newtonian gravity with relativistic velocities (either bulk
or thermal), then there are two relevant UDCs, $G$ and $c$ ($N_{\rm
udc} = 2$), which imply Eq.~(\ref{eq:gravity}) and $\alpha = \eta$,
respectively. Together this implies that $\zeta = \alpha =
\eta$, i.e. that all three scaling coefficients are equal,
\begin{equation}\label{eq:G,c}
(m'_0,\,l'_0,\,t'_0) = \alpha\,(m_0,\, l_0,\, t_0)\ ,\quad\ \ 
Q'(x'_\mu = \alpha x_\mu) = \alpha^{A+B+C} Q(x_\mu)\ .
\end{equation}
\vspace{0.15cm}

\noindent{\bf Magneto-Hydrodynamic (MHD) Simulations:}
the MHD equations are also based on Maxwell's equations, and thus
include $c$ as a UDC, so they require that $\alpha = \eta$. This holds
even in the Newtonian case, where there are two free parameters ($N_f
= 2$) describing the relevant family of physical systems corresponding
to a particular simulation, according to Eq.~(\ref{eq:c}).  If gravity
is included, even if weak or Newtonian gravity, then this introduces a
second UDC, $G$, resulting in only one free parameter describing the
relevant family of physical systems ($N_f = 1$ since $N_{\rm udc} =
2$), according to Eq.~(\ref{eq:G,c}).

\vspace{0.2cm}
\noindent{\bf Simulations with discrete particles:}
there are various types of simulations that aim to describe the
motions of discrete point-like particles, under the influence of the
mutual forces that they exert on each other, rather than a continuous
medium that is described by hydrodynamic or MHD equations.  Here I
briefly go over two important types of such simulations.

The first type is particle in cell (PIC) simulations of the
motions of charged particles of either positive or negative electric
charge under the mutual electromagnetic forces that they exert on each
other. In this case Maxwell's equations introduce $c$ as a UDC
(implying $\alpha=\eta$). If we do not mind that the scaling would
change the rest mass and/or electric charge of particles, then this
would be the only constraint ($N_{\rm udc} = 1$), implying $N_f = 2$
and Eq.~(\ref{eq:c}). If, however, it is important for us to
accurately model a specific particle species (such as
electrons/positrons) of a given universal rest mass and electric
charge, then this would add two more constraints (and altogether
$N_{\rm udc} = 3$), thus removing all the remaining freedom in the
scaling parameters ($N_f = 0$) and implying $\zeta=\alpha=\eta=1$.

The second type is $N$-body simulations, that are often used in
cosmology and stellar or planetary dynamics, where $N$ point-like
masses move under their mutual gravitational forces. Since gravity is
the only force involved, $G$ must obviously remain constant, implying
Eq.~(\ref{eq:gravity}). If there are only Newtonian gravity and
velocities then there are two free parameters ($N_f = 2$, $N_{\rm udc}
= 1$) and Eq.~(\ref{eq:G}) holds. Otherwise, if relativistic effects
cannot be neglected, then $c$ also enters the relevant equations as a
second UDC ($N_{\rm udc} = 2$) resulting in only one free parameter
($N_f = 1$), and implying Eq.~(\ref{eq:G,c}).

In cosmological $N$-body simulations with Newtonian gravity and
velocities, the only bona fide UDC is $G$, implying Eq.~(\ref{eq:G})
with $N_f = 2$. The situation is more complicated, however, since we
usually want the simulations to agree with the cosmological model of
our observed universe, whose parameters are reasonably well
known. Thus, some of these cosmological parameters might be treated as
UDCs, depending on the purpose of the simulations.

For example, if the Hubble constant $H_0$ is treated as a UDC, then it
would imply $\eta = 1$, which together with $G$ (that implies $\zeta
=\alpha^3$; Eq.~[\ref{eq:gravity}]), results in $N_{\rm udc} = 2$,
$N_f = 1$ and
\begin{eqnarray}\nonumber
(m'_0,\,l'_0,\,t'_0) = (\alpha^3 m_0,\, \alpha l_0,\, t_0)\ ,\quad
\\ \label{eq:cosmo}
Q'\left(t' = t,\,\vec{r'} = \alpha\vec{R}\right) =
\alpha^{B+3A} Q(t,\vec{r})\ .
\end{eqnarray}
In particular this would leave the mass density unchanged, $\rho'(t')
=\rho'(t) = \rho(t)$, so that the effective $\Omega_M(t)$ that is
implied by the average value of $\langle\rho\rangle$ over the
computational box would still follow the same original assumed
cosmology.

\begin{table}
\centering
\caption{
The freedom in the choice of the basic physical units for different
types of numerical simulations. The columns, from left to right,
list the type of simulation, the relevant independent (in terms of
their units) universal dimensional constants (UDCs), their number
$N_{\rm udc}$, the number of free parameters $N_f$ they allow for
a rescaling of the units, and the imposed relation between the
rescaling factors for mass ($\zeta$), length ($\alpha$)
and time ($\eta$) units.
\newline
$^\dagger\,$If one or more of the cosmological parameters (such as
$H_0$ or $\sigma_8$) are treated as UDCs this reduces $N_f$ -- see
the discussion in \S~\ref{sec:sim}.}
\label{tab:sum}
\begin{tabular}{|l|c|c|c|c|}
\hline
\multirow{2}{*}{type of simulation} & \multirow{2}{*}{UDCs} & 
\multirow{2}{*}{$N_{\rm udc}$} & \multirow{2}{*}{$N_f$} & constraints \\
 & & & & on rescaling\\
\hline\hline
Newtonian hydrodynamics & --- & 0 & 3 & --- \\ \hline
relativistic hydrodynamics; & \multirow{2}{*}{$c$} & \multirow{2}{*}{1} & \multirow{2}{*}{2} & \multirow{2}{*}{$\alpha = \eta$}\\
Newtonian/relativistic MHD & & & & \\ \hline
Newtonian gravity (e.g. in & \multirow{3}{*}{$G$} & \multirow{3}{*}{1} & \multirow{3}{*}{2} & \multirow{3}{*}{$\zeta = \alpha^3\eta^{-2}$} \\
stellar/planetary dynamics, & & & & \\
cosmological $N$-body$\,^\dagger$) & & & & \\ \hline
general relativistic & \multirow{4}{*}{$G$, $c$} & \multirow{4}{*}{2} & \multirow{4}{*}{1} & \multirow{4}{*}{$\zeta = \alpha = \eta$}\\
hydrodynamics or MHD; & & & & \\ 
Newtonian gravity + MHD & & & & \\ 
or relativistic velocities & & & & \\ \hline
particle in cell (PIC); & $c$ & 1 & 2 & $\alpha=\eta$\\
PIC + particular particles & $c$, $q$, $m$ & 3 & 0 & $\zeta$$\,=\,$$\alpha$$\,=\,$$\eta$$\,=\,$$1$\\
\hline
\end{tabular}
\end{table}

In cosmological $N$-body simulations, the initial conditions are
considered to be scale-invariant, since the amplitude of the initial
fluctuations in the gravitational potential are (at least nearly)
independent of the wavenumber $k$. However, the corresponding
fluctuations in density scale as
$\langle\delta\rho/\rho\rangle(k)\propto k^2$. This introduces a time
dependent scale, $l_1(t) =2\pi/k_1(t)$, at which the density
fluctuations become of order unity,
$\langle\delta\rho/\rho\rangle[k_1(t)]\equiv 1$, and thus enter the
strongly non-linear stage of their evolution. This scale changes under
a rescaling of the length units, and so does $\sigma_8$ or the
normalization of the initial power spectrum (i.e., the length scale
that $\sigma_8$ represents would no longer be $8h^{-1}\;$Mpc upon
rescaling of the length, but instead $\alpha\times 8h^{-1}\;$Mpc).
Sticking to the observed value of $\sigma_8$, and treating both $H_0$
and $\sigma_8$ as UDCs, would effectively remove the last degree of
freedom in our rescaling (i.e. result in $N_f = 0$). If, however, the
cosmological parameters such as $H_0$ or $\sigma_8$ are not treated as
UDCs, and are allowed to vary (even if only over a limited range that
is consistent with current observational constraints), then a
rescaling of the units given by Eq.~(\ref{eq:G}) could help to reduce
the number of simulations required in order to numerically study a
large parameter space with different cosmologies.

\section{\bf Caveats}
\label{sec:caveats}

Depending on the physics that are included in a simulation, further
restrictions may arise in cases where there are additional UDCs or
typical scales.  In the previous section, the {\bf equation of state}
was implicitly assumed not to introduce any UDC, such as in the case
of a simple polytropic equation of state, $p = K\rho^\gamma$, where
$K$ can vary (with the specific entropy). However, this is not always
the case.

For example, {\bf degeneracy pressure} in the Newtonian regime fixes
the value of $p/\rho^{5/3} \sim \hbar^2/(m_d m_{\rm eff}^{5/3})$ where
$m_d$ ($n_d$) is the mass (number density) of the degenerate species
while $m_{\rm eff} = \rho/n_d$ (here it is assumed that $\rho$ is used
as a primary hydrodynamic variable, rather than the number density
$n$). This requires that $\zeta =\alpha^6\eta^{-3}$. Since degeneracy
pressure is usually important only when gravity plays a role as well,
this would also require $\zeta = \alpha^3\eta^{-2}$ or altogether,
$\eta=\alpha^3=\zeta^{-1}$. In the relativistic regime, where the
uncertainty principle implies relativistic velocities of the
degenerate species, $p/\rho^{4/3} \sim
\hbar c/m_{\rm eff}^{4/3}$ is fixed, implying $\zeta
=\alpha^9\eta^{-6}$. Together with gravity that introduces $G$, this
implies $\zeta = 1$ and $\eta=\alpha^{3/2}$. For example, the
Chandrasekhar mass is approximately given by the $3/2$ power of the
ratio of these two constants, $M_{\rm Ch} \sim (\hbar c/m_{\rm
eff}^{4/3}G)^{3/2}= M_{\rm Planck}^3/m_{\rm eff}^2$.  The transition
between the two regimes of degeneracy pressure occurs when the mean
distance between degenerate particles is comparable to their Compton
wavelength, thus fixing an absolute length-scale in the problem and
requiring $\alpha = 1$ if it appears in the simulation. Together with
gravity this would leave no free parameter ($N_f = 0$ since $N_{\rm
udc} = 3$), and require $\zeta =\alpha = \eta = 1$.

The {\bf optical depth} $\tau$ determines the probability for
interaction, $1-e^{-\tau}$, and must therefore remain unchanged. Thus,
if $\rho$ (rather than $n$) is a primary hydrodynamic variable then
since $d\tau = \rho\kappa_*dl$, once the opacity coefficient
$\kappa_*$ of the matter is specified it should not change, and since
it has units of $l^2/m$, this implies $\zeta = \alpha^2$. With the
inclusion of radiation that introduces $c$ as a UDC and requires
$\alpha = \eta$, this implies $\zeta = \alpha^2 =
\eta^2$. If, alternatively, $n$ is the primary hydrodynamic variable
then since $d\tau = \sigma_* n dl$ this requires the cross-section
$\sigma_*$ not to change and thus $\alpha = 1$, which together
with the inclusion of radiation (implying $\eta = \alpha$) gives
$\alpha = \eta = 1$. If the mass of each particle is also to remain
constant, this requires $\zeta = 1$ leaving no degree of freedom and
implying $\zeta=\alpha=\eta=1$.

{\bf Optically thick} radiation, or {\bf radiation pressure} can also
introduce UDCs.  A black body, e.g., emits a power per unit area of
$\sigma T^4$ and has a pressure of $p_{\rm rad} =
\frac{1}{3}aT^4$, thus introducing the Stefan-Boltzmann constant
$\sigma = ac/4$ and the radiation constant $a$. Their ratio introduces
$c = 4\sigma/a$ (the radiation streaming velocity) that implies
$\alpha = \eta$. Since $k_{\rm B}T \approx p_{\rm rad}/n$ then $p_{\rm
rad}/(k_{\rm B}T)^4$ that has unit of $(ml^3t^{-2})^{-3}$ must also
remain the same, implying $\zeta =\eta^2\alpha^{-3}$, and together
with the previous constraint, $\alpha = \eta =
\zeta^{-1}$. If gravity is added as well then no freedom is left ($N_f
= 0$ and $\zeta = \alpha = \eta = 1$).

{\bf Radiation reaction} (the force on accelerating charged particles
due to the back-reaction to the radiation they emit) or the effects of
{\bf radiative losses} on the cooling of the radiating particles, can
introduce additional dimensional parameters, that are universal for a
given particle species, such as electrons, and thus introduce
constraints on the scaling parameters.

\begin{table}
\caption{
The dimensional-based scalings for the GRB afterglow synchrotron
spectrum, in terms of $\zeta$ and $\alpha = \eta$ (in column 3) or
$\kappa$ and $\lambda$ (in column 4), for the flux density within the
different power-law segments (PLSs, $Q\to
F_{\rm\nu,A}$--$F_{\rm\nu,G}$; top part), the spectral break
frequencies ($Q\to\nu_1$--$\nu_{11}$; middle part), and the flux
density at the break frequencies ($Q\to F_{\nu,1}$--$F_{\nu,11}$;
bottom part). The notation for the different PLSs and break
frequencies follow \citet{GS02}. Column 2 gives the dependence of
$F_\nu$ on $\nu$ in each PLS for $F_{\rm\nu,A}$--$F_{\rm\nu,G}$, and
otherwise the relevant break frequencies.}
\label{tab:PLS}
\begin{tabular}{|c|c|c|c|}
\hline
$Q$ & $\nu$ & $\zeta,\,\alpha$ & $\kappa,\,\lambda$ \\
\hline\hline
$F_{\rm\nu,A}$ & $\nu^{5/2}$     & $\zeta^{-1/4}\alpha^{11/4}$         & $\kappa^{2/3}\lambda^{-11/12}$ \\
$F_{\rm\nu,B}$ & $\nu^{2}$       & $\zeta^{0}\alpha^{2}$               & $\kappa^{2/3}\lambda^{-2/3}$ \\
$F_{\rm\nu,C}$ & $\nu^{11/8}$    & $\zeta^{1/8}\alpha^{13/8}$          & $\kappa^{2/3}\lambda^{-13/24}$ \\
$F_{\rm\nu,D}$ & $\nu^{1/3}$     & $\zeta^{4/3}\alpha^{-1}$            & $\kappa^{1}\lambda^{1/3}$ \\
$F_{\rm\nu,E}$ & $\nu^{1/3}$     & $\zeta^{2}\alpha^{-7/3}$            & $\kappa^{11/9}\lambda^{7/9}$ \\
$F_{\rm\nu,F}$ & $\nu^{-1/2}$    & $\zeta^{3/4}\alpha^{-1/4}$          & $\kappa^{2/3}\lambda^{1/12}$ \\
$F_{\rm\nu,G}$ & $\nu^{(1-p)/2}$ & $\zeta^{(p+5)/4}\alpha^{-3(p+1)/4}$ & $\kappa^{1}\lambda^{(p+1)/4}$ \\
$F_{\rm\nu,H}$ & $\nu^{-p/2}$    & $\zeta^{(p+2)/4}\alpha^{(2-3p)/4}$  & $\kappa^{2/3}\lambda^{(3p-2)/12}$ \\
\hline
$\nu_m$ & $\nu_2,\nu_4,\nu_9$ & $\zeta^{1/2}\alpha^{-3/2}$ & $\kappa^{0}\lambda^{1/2}$ \\
$\nu_c$ & $\nu_3,\nu_{11}$      & $\zeta^{-3/2}\alpha^{5/2}$ & $\kappa^{-2/3}\lambda^{-5/6}$ \\
$\nu_{\rm ac}$ & $\nu_7$          & $\zeta^{1/5}\alpha^{-3/5}$ & $\kappa^{0}\lambda^{1/5}$ \\
$\nu_{\rm sa}$ & $\nu_1$          & $\zeta^{4/5}\alpha^{-9/5}$ & $\kappa^{1/5}\lambda^{3/5}$ \\
$\nu_{\rm sa}$ & $\nu_5$          & $\zeta^{\frac{6+p}{8+2p}}\alpha^{-\frac{14+3p}{8+2p}}$ & $\kappa^{\frac{2}{12+3p}}\lambda^{\frac{14+3p}{24+6p}}$ \\
$\nu_{\rm sa}$ & $\nu_6$          & $\zeta^{\frac{3+p}{10+2p}}\alpha^{-\frac{9+3p}{10+2p}}$ & $\kappa^{0}\lambda^{\frac{3+p}{10+2p}}$ \\
$\nu_{\rm sa}$ & $\nu_8$          & $\zeta^{1/3}\alpha^{-1}$ & $\kappa^{0}\lambda^{1/3}$ \\
$\nu_{\rm sa}$ & $\nu_{10}$       & $\zeta^{9/5}\alpha^{-19/5}$  & $\kappa^{8/15}\lambda^{19/15}$ \\
\hline
$F_{\nu,1}$
         & $\nu_1$            & $\zeta^{8/5}\alpha^{-8/5}$ & $\kappa^{16/15}\lambda^{8/15}$ \\
$F_{\nu,{\rm max}}$ & $\nu_2,\nu_{11}$ & $\zeta^{3/2}\alpha^{-3/2}$ & $\kappa^{1}\lambda^{1/2}$ \\
$F_{\nu,3}$  & $\nu_3$    & $\zeta^{(2p+1)/2}\alpha^{(1-4p)/2}$ & $\kappa^{(p+2)/3}\lambda^{(4p-1)/6}$ \\
$F_{\nu,4}$  & $\nu_4$    & $\zeta^{1}\alpha^{-1}$ & $\kappa^{2/3}\lambda^{1/3}$ \\
$F_{\nu,5}$  & $\nu_5$    & $\zeta^{\frac{13+2p}{8+2p}}\alpha^{-\frac{13+2p}{8+2p}}$ & $\kappa^{\frac{13+2p}{12+3p}}\lambda^{\frac{13+2p}{24+6p}}$ \\
$F_{\nu,6}$  & $\nu_6$    & $\zeta^{\frac{5+2p}{10+2p}}\alpha^{\frac{5-2p}{10+2p}}$ & $\kappa^{2/3}\lambda^{\frac{2p-5}{30+6p}}$ \\
$F_{\nu,7}$  & $\nu_7$    & $\zeta^{2/5}\alpha^{4/5}$ & $\kappa^{2/3}\lambda^{-4/15}$ \\
$F_{\nu,8}$  & $\nu_8$    & $\zeta^{7/12}\alpha^{1/4}$ & $\kappa^{2/3}\lambda^{-1/12}$ \\
$F_{\nu,9}$  & $\nu_9$    & $\zeta^{1/2}\alpha^{1/2}$ & $\kappa^{2/3}\lambda^{-1/6}$ \\
$F_{\nu,10}$ & $\nu_{10}$ & $\zeta^{13/5}\alpha^{-18/5}$ & $\kappa^{7/5}\lambda^{6/5}$ \\
\hline
\end{tabular}
\end{table}

\section{GRB afterglows}
\label{sec:afterlow}

The dynamics of GRB jets during the afterglow stage have been
numerically modeled using special relativistic hydrodynamic
simulations (\citealt{Granot01}; \citealt{CGV04}; \citealt{ZM09}; 
Mimica, Giannios \& Aloy 2009, 2010; \citealt{vanEerten10}; 
\citealt{MK10}; \citealt{WWF11}; De Colle et al. 2011a, 2011b).  
As discussed above, for such simulations there is one UDC, $c$, which
implies $\alpha = \eta$ and Eq.~(\ref{eq:c}). It has recently been
pointed out\footnote{\citet{Scheck02} have outlined a similar scaling
in a different context.} \citep{vE11a} that the dynamics in this case
obey a simple scaling relation,
\begin{equation}
\frac{E'}{E} = \kappa\ ,\quad\quad 
\frac{\rho'}{\rho} = \lambda\ ,\quad\quad
\frac{l'}{l} = \frac{t'}{t} = \fracb{\kappa}{\lambda}^{1/3}\ ,
\end{equation}
which was justified by resorting to dimensionless or similarity
variables. However, this scaling simply arises from the freedom in the
choice of the basic physical units, as described above.  In
particular, it corresponds to $\zeta =\kappa$ and $\alpha = \eta =
(\kappa/\lambda)^{1/3}$, or equivalently to $\kappa = \zeta$ and
$\lambda = \zeta/\alpha^3 =\zeta/\eta^3$.  This scaling holds
regardless of the initial conditions or symmetry of the problem, and
has nothing to do with self-similarity of the hydrodynamics.

It was also pointed out recently \citep{vE11b} that this scaling of
the dynamics\footnote{There they use the scaling $n'/n =\lambda$ for
the number density, but this is effectively equivalent to $\rho'/\rho
=\lambda$ since they assume that $\rho/n = m_p = {\rm const}$.} can
also be extended to a similar scaling of the resulting afterglow
synchrotron emission or the observed flux density $F_\nu$, within each
power-law segment (PLS) of the spectrum. This arises since within each
PLS the local emissivity can be expressed as the product of a
dimensional constant and a dimensionless function of the hydrodynamic
quantities, so that a change in the basic units would affect only the
dimensional constant, which would scale in a simple
way.\footnote{According to the units of the part that scales with the
hydrodynamics variables, and does not involve the distance from the
source to the observer or UDCs such as the electron or proton mass or
electric charge.}  Therefore, such a rescaling of the basic physical
units holds quite generally within each PLS, regardless of the
dynamics. In particular, the same rescaling holds in the early
relativistic \citep{BM76} and late Newtonian \citep{Sedov46,Taylor50}
(quasi-) spherical self-similar phases, as well as in the intermediate
phase where the dynamics are not self-similar. Moreover, this scaling
depends only on the PLS, and within a given PLS it does not depend on
the external density profile (and would be the same for a uniform
external medium and for a wind-like external medium).

It has been demonstrated that when the dynamics are self-similar, a
more elaborate scaling exists in which the flux density $F_\nu$ within
each PLS scales as a power-law with essentially all of the model
parameters \citep{SPN98,GS02,vE11b}. However, the dependences on the
individual model parameters change between the relativistic
\citet{BM76} and the Newtonian Sedov-Taylor self-similar regimes, and
such simple power-law dependences on all of the model parameters do
not exist in the intermediate phase, or whenever the dynamics are not
self-similar. 

The dimensional-based scalings of $F_\nu$ hold only locally within
each PLS, and change between different PLSs. This corresponds to
different scalings for the break frequencies that separate the PLSs,
so that their ratios changes under such a scaling, despite being
dimensionless. The lack of a global rescaling of the units for the
observed radiation results in the need to parameterize and change ``by
hand'' the spectral regime in order to calculate the lightcurve at a
given observed frequency as it switches between different PLSs (when
it is crossed by a break frequency).

The lack of such a global rescaling of the units for $F_\nu$ can be
understood as follows. Technically, it can be attributed to the fact
that the local emissivity is separable, i.e. can be expressed as a
product of a dimensional constant and a dimensionless function of the
hydrodynamic variables, only within each PLS, and that this
dimensional constant that determines the scaling changes between
different PLSs. The more basic reason behind this is that the emission
process introduces additional UDCs relative to the dynamics, even in
the optically thin regime.  For example, the radiation cares also
about the total number of particles, i.e. about the number density $n$
and not only about the rest-mass density $\rho$, while $\rho/n\equiv
m_{\rm eff}$ is usually taken to be constant (often set to the proton
mass, $m_p$), and thus introduces a new UDC. Additional UDCs are
introduced, e.g., through the synchrotron break frequencies, since
they relate to the typical synchrotron frequency and cooling of the
radiating relativistic electrons, which have a universal mass and
electric charge.

Let us consider an emitting region of bulk Lorentz factor $\Gamma$, in
the downstream region of a shock with a relative upstream to
downstream Lorentz factor $\Gamma_{\rm ud}$ and upstream proper rest
mass density $\rho_{\rm u}$. For the afterglow forward shock
$\rho_{\rm u}$ is the external density and $\Gamma_{\rm ud} - 1
\approx \Gamma_{\rm ud} = \Gamma \gg 1$, while for the reverse shock
$\rho_{\rm u}$ is the density of the original outflow and typically
$\Gamma\gg\Gamma_{\rm ud} > \Gamma_{\rm ud}-1 \sim 1$. Thus, both
shocks can be treated together. The comoving magnetic field scales as
$B^2\propto \epsilon_B\Gamma_{\rm ud}(\Gamma_{\rm ud}-1)\rho_{\rm u}$,
and the typical Lorentz factor of the electron random motions scales
as $\gamma_m \sim \epsilon_e(m_p/m_e)(\Gamma_{\rm ud}-1) \propto
\Gamma_{\rm ud}-1$. Thus, the typical synchrotron frequency scales as
$\nu_m \sim \Gamma(eB/m_ec)\gamma_m^2 \propto 
\epsilon_B^{1/2}\epsilon_e^2\Gamma\Gamma_{\rm ud}^{1/2}
(\Gamma_{\rm ud}-1)^{5/2}\rho_{\rm u}^{1/2}\propto \rho_{\rm u}^{1/2}$
so that $\nu'_m/\nu_m = \zeta^{1/2}\alpha^{-3/2} \to \lambda^{1/2}$
since $\gamma_m$, $\Gamma$, $\Gamma_{\rm ud}$, as well as the shock
microphysics parameters $\epsilon_e$ and $\epsilon_B$ are all
invariant under rescalings of the basic physical units that conserve
$c$ ($\alpha =\eta$).  Note that the part involving UDCs, $e/m_e c$,
was not included in the scaling, since it is universal and does not
change with the scaling of the hydrodynamic variables. The cooling
break frequency scales as $\nu_c\sim\Gamma(eB/m_ec)\gamma_c^2 \propto
\Gamma^{-1}B^{-3}t_{\rm obs}^{-2}\propto
\Gamma^{-1}[\epsilon_B\Gamma_{\rm ud}(\Gamma_{\rm ud}-1)\rho_{\rm u}]^{-3/2}
t_{\rm obs}^{-2} \propto \rho_{\rm u}^{-3/2}t_{\rm obs}^{-2}$, 
and thus $\nu'_c/\nu_c =\zeta^{-3/2}\alpha^{9/2}\eta^{-2}\to
\zeta^{-3/2}\alpha^{5/2} \to\kappa^{-2/3}\lambda^{-5/6}$,
where $t_{\rm obs}$ is the observed time (when the emitted photons
reach the observer) and $\gamma_c = 6\pi m_e c/(\sigma_T B^2\Gamma
t_{\rm obs}) \propto B^{-2}\Gamma^{-1}t_{\rm obs}^{-2}$ is the random
Lorentz factor to which the electrons cool on the dynamical time.
Again, parts involving UDCs, such as $e/m_e c$ or $m_e c/\sigma_T$,
were not included in the scaling. A global rescaling of the units
would require $\nu_m\rho_{\rm u}^{-1/2}$ and $\nu_c\rho_{\rm
u}^{3/2}t_{\rm obs}^2$ with units of $m^{-1/2}l^{3/2}t^{-1}$ and
$m^{3/2}l^{-9/2}t$, respectively, to remain invariant, thus implying
$\zeta =\alpha^3\eta^{-2}\to\alpha$ and $\zeta =\alpha^3\eta^{-2/3}\to
\alpha^{7/3}$, or altogether $\zeta = \alpha = \eta = 1$, which 
eliminates all of the freedom in such a rescaling. The peak synchrotron
flux density scales as $F_{\nu,{\rm max}} \propto \Gamma B N_e \propto
\epsilon_B^{1/2}\Gamma\Gamma_{\rm ud}^{1/2}(\Gamma_{\rm ud}-1)^{1/2}
\rho_{\rm u}^{1/2}M \propto \rho_{\rm u}^{1/2}M$ (where $N_e$ and $M$ 
are, respectively, the isotropic equivalent number of emitting
electrons and rest mass in the shocked region, and $M/N_e = m_{\rm
eff} = {\rm const}$), which implies that 
$F'_{\nu,{\rm max}}/F_{\nu,{\rm max}} =
\zeta^{3/2}\alpha^{-3/2}\to\kappa\lambda^{1/2}$. Note that the distance 
to the observer, $D$, is not included in the scaling of $F_{\nu,{\rm
max}}$ since it does not change with the hydrodynamic variables.

Even though there is no non-trivial global scaling of the units that obeys
Eq.~(\ref{eq:scaling}) for the flux density ($Q \to F_\nu$), such a
scaling still works locally within each PLS (labeled by a subscript
`$i$'),
\begin{equation}
F'_{\nu,i}(t'_{\rm obs}=\alpha t_{\rm obs}) = 
\zeta^{a_i}\alpha^{b_i}F_{\nu,i}(t_{\rm obs})\ ,
\end{equation}
where the dependence on $t$ and $\vec{r}$ is replaced by $t_{\rm
obs}$. This can be understood since the flux density within each PLS
is the product of $F_{\nu,{\rm max}}$ and certain fixed powers of the
break frequencies ($\nu_m$, $\nu_c$, and the self-absorption frequency
that is not discussed here for simplicity), whose scalings can be
derived from simple dimensional considerations (as shown above). For
example, $F_{\rm\nu,D} \approx F_{\nu,{\rm max}}(\nu/\nu_m)^{1/3}$ and
$F_{\rm\nu,F}\approx F_{\nu,{\rm max}} (\nu/\nu_c)^{-1/2}$ for PLSs D
and F, respectively, using the notations of \citet{GS02}. This implies
that $F'_{\nu,D}/F_{\nu,D}
=\zeta^{4/3}\alpha^{-1}\to\kappa\lambda^{1/3}$ ($a_D = 4/3$ and $b_D =
-1$) and $F'_{\nu,F}/F_{\nu,F}
=\zeta^{3/4}\alpha^{-1/4}\to\kappa^{2/3}\lambda^{1/12}$ ($a_F = 3/4$
and $b_F = -1/4$). As illustrative examples of how the scalings for
self-absorbed PLSs may be derived, one can readily obtain that
$F_{\rm\nu,B} \approx \pi(R/\Gamma D)^2(2\nu^2/c^2)\Gamma\gamma_m m_e
c^2 \propto \nu^2R^2$ implying $F'_{\nu,B}/F_{\nu,B}
=\zeta^{0}\alpha^{2}\to\kappa^{2/3}\lambda^{-2/3}$ ($a_B = 0$ and $b_B
= 2$), while for PLS A $\gamma_m$ is replaced by $\gamma_e(\nu)
\propto (\nu/\Gamma B)^{1/2}$ [obtained from requiring $\nu\sim \nu_{\rm
syn}(\gamma_e) \sim \Gamma(eB/m_e c)\gamma_e^2$], implying
$F_{\rm\nu,A} \propto \nu^{5/2}R^2\rho_{\rm u}^{-1/4}$ and
$F'_{\nu,A}/F_{\nu,A}
=\zeta^{-1/4}\alpha^{11/4}\to\kappa^{2/3}\lambda^{-11/12}$ ($a_A =
-1/4$ and $b_A = 11/4$).  Therefore, these scalings (or $a_i$ and
$b_i$) do not depend on the external density profile or on the details
of the dynamics (and are the same in the relativistic and Newtonian
self-similar regimes, when the dynamics are not self-similar, or for
the reverse shock). All of the different scalings are summarized in
table~\ref{tab:PLS}.

\section{\bf Discussion}
\label{sec:dis}

The freedom in the choice of units in the dynamical equations that
describe the evolution of different types of physical systems and in
their solutions, has been outlined and elucidated. The main results
are summarized in Table~\ref{tab:sum}. While the emphasis was on
numerical solutions of the dynamical equations through simulations,
similar scalings hold equally well for analytic solutions of the same
equations. The number of free parameters $N_{\rm f}$ that describe the
family of physical systems that corresponds to a given solution of
such a set of equations is given by $\max(0,3-N_{\rm udc})$
(Eq.~[\ref{eq:N_f}]), where $N_{\rm udc}$ is the number of independent
(in terms of their units) universal dimensional constants (UDCs, such
as $c$, $G$, $\hbar$, $m_e$, etc.). This corresponds to the three
basic physical units (of mass, length and time) while accounting for
the independent constraints on their possible rescalings. Such
rescalings of the basic units are potentially relevant to many
different areas of research, such as plasma physics, astrophysics,
cosmology, fluid dynamics or Earth and planetary sciences. They can
prove very useful in numerical studies of various physical systems,
and save precious computational resources, especially in systematic
numerical studies of a large parameter space.

\vspace{0.2cm}
The author thanks E. Nakar, F. van den Bosch, F. De Colle, T. Piran,
O. Bromberg, E. Ramirez-Ruiz and the anonymous referee for useful
discussions, suggestions or comments on the manuscript.  This research
was supported by the ERC advanced research grant ``GRBs''.

\label{lastpage}


\begin{thebibliography}{}

\bibitem[Amit et al.(2010)]{Amit10}
Amit, H., Leonhardt, R., \& Wicht, J. 2010, SSRv, 155, 293

\bibitem[Blandford \& McKee(1976)]{BM76}
Blandford, R.~D., \& McKee, C.~F. 1976, Phys. Fluids, 19, 1130

\bibitem[Birn et al.(2001)]{Birn01}
Birn, J., et al. 2001, Journal of Geophysical Research, 106, 3715

\bibitem[Cannizzo et al.(2004)]{CGV04}
Cannizzo, J.K., Gehrels, N., \& Vishniac, E.T. 2004, ApJ, 601, 380

\bibitem[De Colle et al.(2011a)]{DC11a}
De Colle, F., Granot, J.,  Lopez-C\'amera, D., \& Ramirez-Ruiz, E.
2011a, accepted to ApJ (arXiv:1111.6890) 

\bibitem[De Colle et al.(2011b)]{DC11b}
De Colle, F., Ramirez-Ruiz, E., Granot, J.,  \& Lopez-C\'amera, D. 
2011b, submitted to ApJ (arXiv:1111.6667)

\bibitem[Ellis(1968)]{Ellis68}
Ellis, B. 1968, Basic Concepts of Measurement (Cambridge University
Press)

\bibitem[Fendt \& Memola(2008)]{FM08}
Fendt, C., \& Memola, E. 2008, IJMPD, 17, 1677

\bibitem[Gnedin et al.(2004)]{Gnedin04}
Gnedin, O. Y., et al. 2004, ApJ, 616, 16	

\bibitem[Granot et al.(2001)]{Granot01}
Granot, J., Miller, M., Piran, T., Suen, W. M., \& Hughes, P. A. 2001,
in ``GRBs in the Afterglow Era'', ed. E. Costa, F. Frontera, \& J. Hjorth 
(Berlin: Springer), 312

\bibitem[Granot \& Sari(2002)]{GS02}
Granot, J., \& Sari, R. 2002, ApJ, 568, 820

\bibitem[Ishihara et al.(2009)]{Ishihara09}
Ishihara, T., Gotoh, T., \& Kaneda, Y. 2009, AnRFM, 41, 165

\bibitem[Katz, Weinberg \& Hernquist(1996)]{KWH96}
Katz, N., Weinberg, D. H., \& Hernquist, L. 1996, ApJS, 105, 19	

\bibitem[Kravtsov et al.(1997)]{KKK97}
Kravtsov, A. V., Klypin, A. A., \& Khokhlov, A. M. 
1997, ApJS, 111, 73	

\bibitem[Lawson \& Barakos(2011)]{LB11}
Lawson, S. J., \& Barakos, G. N. 2011, PrAeS, 47, 186

\bibitem[Li et al.(2010)]{Li10}
Li, X.-L., et al. 2010, Acta Mechanica Sinica, 26, 795

\bibitem[Meliani \& Keppens(2010)]{MK10} 
Meliani, Z., \& Keppens, R. 2010, A\&A, 520, L3

\bibitem[Mimica, Giannios \& Aloy(2009)]{MGA09}
Mimica, P., D. Giannios, D., \& Aloy, M. A. 2009, A\&A, 494, 879

\bibitem[Mimica, Giannios \& Aloy(2010)]{MGA10}
Mimica, P., D. Giannios, D., \& Aloy, M. A. 2010, IJMPD, 19, 985

\bibitem[Navarro, Frenk \& White(1997)]{NFW97}
Navarro, J.$\,$F., Frenk, C.$\,$S., \& White, S.$\,$D.$\,$M. 1997, ApJ, 490, 493

\bibitem[Pukhov \& Meyer-ter-Vehn(2002)]{PM02}
Pukhov, A., \& Meyer-ter-Vehn, J. 2002, ApPhB, 74, 355	

\bibitem[Sari, Piran \& Narayan(1998)]{SPN98}
Sari, R., Piran, T., \& Narayan, R. 1998, ApJ, 497, L17


\bibitem[Scheck et al.(2002)]{Scheck02}
Scheck, L., Aloy, M. A., Martí, J. M., Gómez, J. L., \& Müller, E., 2002, MNRAS, 331, 615

\bibitem[Sedov(1946)]{Sedov46}
Sedov, L. I. 1946, Prikl. Math. Mekh. 10, 241, no. 2

\bibitem[Spitkovsky(2008)]{Spit08}
Spitkovsky, A. 2008, ApJ, 682, L5	

\bibitem[Springel(2005)]{Spr05}
Springel, V. 2005, MNRAS, 364, 1105	

\bibitem[Springel et al.(2005)]{Springel05}
Springel, V., et al. 2005, Nature, 435, 629	

\bibitem[Taylor(1950)]{Taylor50}
Taylor, G. I. 1950, Proc. R. Soc. London A, 201, 159

\bibitem[van Eerten et al.(2010)]{vanEerten10}
van Eerten, H. J., Zhang, W., \& MacFadyen, A. 2010, ApJ, 722, 235

\bibitem[van Eerten et al.(2011)]{vanEerten11}
van Eerten, H. J., Meliani, Z., Wijers, R. A. M. J., \& Keppens, R.
2011, MNRAS, 410, 2016

\bibitem[van Eerten, van der Horst \& MacFadyen(2011)]{vE11a}
van Eerten, H. J., van der Horst, A. J., \& MacFadyen, A. I. 2011, arXiv:1110.5089

\bibitem[van Eerten \& MacFadyen(2011)]{vE11b}
van Eerten, H. J., \& MacFadyen, A. I. 2011, arXiv:1111.3355

\bibitem[Wallace(2011)]{Wallace11}	
Wallace, J. M. 2009, Phys. Fluids, 21, 1301

\bibitem[Wygoda, Waxman \& Frail(2011)]{WWF11}
Wygoda, N., Waxman, E., \& Frail, D. A. 2011, ApJ, 738, L23

\bibitem[Zhang \& MacFadyen(2009)]{ZM09}
Zhang, W., \& MacFadyen, A. I. 2009, 698, 1261


\end{thebibliography}
\end{document}